\documentclass{article}

\title{\large Graph Isomorphism is PSPACE-complete}  

\author{Matthew Delacorte}
\date{August 30, 2007}

\begin{document}             

\maketitle 
                  
\footnotesize Combining the the results of A.R. Meyer and L.J. Stockmeyer "The Equivalence
Problem for Regular Expressions with Squaring Requires
Exponential Space", and K.S. Booth  "Isomorphism testing for graphs, semigroups, and finite automata are
polynomiamlly equivalent problems" shows that graph isomorphism is PSPACE-complete.

\section {Proof}      
The equivalence problem for regular expressions was shown to be PSPACE-complete by (Meyer and Stockmeyer [2]).  Booth [1] has shown that isomorphism of finite automata is equivalent to graph isomorphism.  Taking these two results together with the equivalence of regular expressions, right-linear grammars, and finite automata see [3] for example, shows that graph isomorphism is PSPACE-complete.


\begin{thebibliography}{99}



\bibitem{Booth}
Booth, K.S. Isomorphism testing for graphs, semigroups, and finite automata are
polynomiamlly equivalent problems, SIAM J. Comput. 7, No 3, (1978), 273-279.


\bibitem{Hopcroft}
Hopcroft, J.E., and Ullman, J.D. (1979), Introduction to Automata Theory, Languages and Computation, Addison-Wesley, Reading, MA. 



\bibitem{Meyer}
Meyer, A.R. and Stockmeyer, L.J. The Equivalence
Problem for Regular Expressions with Squaring Requires
Exponential Space, 13th Annual IEEE Symp. on
Switching and Automata Theory, Oct., 1972,125-129.


\end{thebibliography}
\end{document}